\documentclass[12pt]{iopart_mod}
 
\usepackage{graphicx}	
\usepackage{color}		

\newcommand{\ii}{\dot{\imath}}

\newcommand{\ket}[1]{\vert{#1}\rangle}

\begin{document}

\title{Randomized benchmarking of atomic qubits in an optical lattice}

\author{S. Olmschenk, R. Chicireanu, K. D. Nelson, and J. V. Porto}

\address{Joint Quantum Institute, National Institute of Standards and Technology and University of Maryland Department of Physics, Gaithersburg, MD 20899, USA}
\ead{steven.olmschenk@gmail.com}

\date{\today}

\begin{abstract}
We perform randomized benchmarking on neutral atomic quantum bits (qubits) confined in an optical lattice.  Single qubit gates are implemented using microwaves, resulting in a measured error per randomized computational gate of $1.4(1) \times 10^{-4}$ that is dominated by the system $T_2$ relaxation time.  The results demonstrate the robustness of the system, and its viability for more advanced quantum information protocols.
\end{abstract}

\pacs{03.67.Lx, 37.10.Jk}
\maketitle

\section{Introduction}
Quantum information processing has the potential to revolutionize computation and communication.  However, the physical realization of a quantum information processor places stringent demands on the efficiency and reliability of the individual quantum gate operations that drive the computation~\cite{ladd:qc_expt_review}.  Although many methods have been developed to characterize the performance of quantum gates, these may be difficult or computationally intensive to implement, and may be unable to distinguish between initialization, read-out, and gate errors~\cite{kuah:prep_qpt, mohseni:qpt}.

Randomized benchmarking has been proposed as a way to efficiently evaluate the performance of a quantum gate~\cite{knill:benchmarking}.  The basic protocol is to apply a random sequence of unitary operations, and measure the decay in the average output state fidelity as a function of the number of applied gates.  This procedure has a number of advantages over other evaluation methods.  The randomization of the sequence depolarizes the noise in the system, and ensures that the aggregate result is independent of a particular gate or proper subset of gates.  Moreover, randomized benchmarking distinguishes between the error of the quantum gate and errors due to initialization and state detection.  Finally, this protocol determines the average fidelity of the gate at an arbitrary point in a sequence of operations, establishing independence of the error rate to changes in the experiment during the course of the quantum computation.  Randomized benchmarking has recently been used to evaluate quantum gates in trapped atomic ion~\cite{knill:benchmarking, biercuk:q_control}, liquid-state NMR~\cite{ryan:nmr_benchmarking}, and solid-state qubits~\cite{chow:solid-state_benchmarking, chow:optimal_control}.

We implement randomized benchmarking on ultracold atomic qubits trapped in an optical lattice.  The single qubit gates are driven by microwave pulses with precisely controlled frequency, duration and phase.  The measured error per randomized computational gate of $1.4(1) \times 10^{-4}$ is dominated by decoherence of the trapped atomic spins, and is indicative of excellent control over the applied gate operations.  We expect that straight-forward upgrades to the system will further improve both the coherence of the qubits and the fidelity of the quantum gates.

\section{Experimental Setup}
We prepare an ultracold cloud of rubidium atoms (${}^{87}$Rb) by first passing atoms from a hot sample through a Zeeman slower~\cite{phillips:zeeman_slower}, and then laser-cooling them in a six-beam magneto-optical trap~\cite{raab:mot}.  The atoms are then confined by a Ioffe-Pritchard-style magnetic trap, and are cooled by forced radiofrequency (rf) evaporation~\cite{pritchard:rf_evap}.  Subsequently, a 1.5 $\mu$m dipole-force trap and quadrupole magnetic field retain the atoms, where the addition of the quadrupole magnetic field assists in supporting the atoms against gravity and allows for confinement along the propagation direction of the dipole beam to be comparable to that provided by the beam in the transverse directions.  Here the atoms are further evaporatively cooled by gradually decreasing the intensity of the 1.5 $\mu$m light, resulting in a spin-polarized Bose-Einstein condensate~\cite{lin:quad_dipole_bec}.

Transitions between the ground level hyperfine manifolds are driven by applying microwave radiation at frequencies near 6.8 GHz.  The required frequencies are generated by mixing a low frequency (0 to 180 MHz) output from either a direct-digital synthesizer (DDS) or an arbitrary waveform generator (AWG) with the frequency-doubled output of a microwave signal generator.\footnote{Rhode \& Schwartz SMT 06.}  The use of a DDS or AWG enables simple control of the frequency and phase of the radiation; here, the AWG is used for efficient frequency sweeps.\footnote{Agilent 33250A Function/Arbitrary Waveform Generator.}  The mixer output passes through a TTL-controlled rf switch (producing approximately square-shaped pulses), then is amplified and sent to a microwave horn directed at the trapped atoms.

Before loading the atoms into the optical lattice, we reduce the total number of trapped atoms by sweeping the frequency of applied microwave radiation near 6.8 GHz to transfer a fraction of the atoms from the trapped ${}^2 S_{1/2}$$\ket{F = 1,m_F = -1}$ $\equiv \ket{1,-1}$ state to the anti-trapped ${}^2 S_{1/2}$$\ket{F = 2,m_F = -2}$ $\equiv \ket{2,-2}$ state.  Here, $F$ is the total angular momentum of an atom, and $m_F$ is its projection along a quantization axis defined by an external magnetic field.  Using this microwave sweep technique, we are able to prepare an ultracold atomic sample of approximately $2 \times 10^4$ atoms.

The atoms are adiabatically loaded into a three dimensional, red-detuned ($\lambda_l = 810.5$ nm) optical lattice by exponentially increasing the intensity of the lattice beams to a maximum value over a duration of about 200 ms.  In two directions, retro-reflection of an intensity-stabilized beam results in a lattice with spacing between sites of $\lambda_{l}/2 = 405$ nm; the optical configuration is similar to the one described in Ref.~\cite{sebby-strabley:double-well_lattice}.  The lattice in the third dimension is generated by two beams (not intensity-stabilized) propagating at an angle of about $160^\circ$, resulting in a lattice spacing of approximately 411 nm.  The depth of the potential between adjacent lattice sites is measured to be about $30 E_r$, determined by pulsing the lattice beams and observing an atom diffraction pattern in time-of-flight~\cite{ovchinnikov:atom_diffraction}.  Here the recoil energy is defined by $E_r/h = h/(2 \lambda_l^2 m) \approx 3.5$ kHz, where $h$ is Planck's constant and $m$ is the mass of the confined atom.  The low initial atom number and measured lattice depth results in no more than one atom per lattice site, as confirmed by microwave spectroscopy~\cite{campbell:mott_shells}.

\begin{figure}
	\centering
	\includegraphics[width=1.0\columnwidth,keepaspectratio]{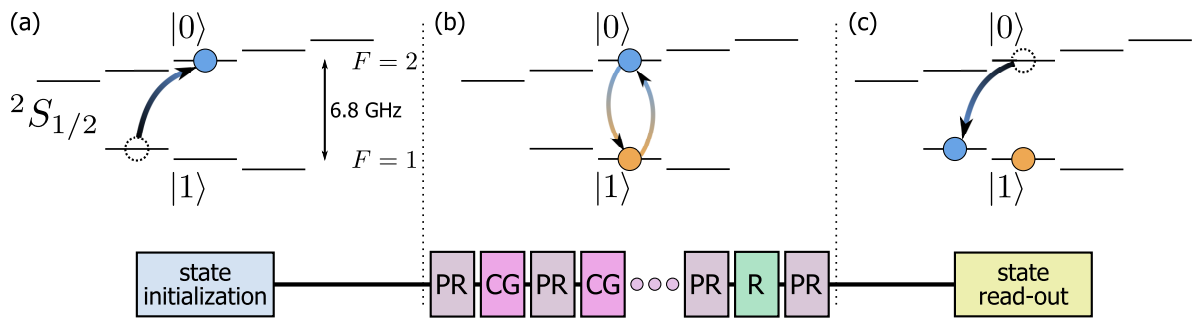}
	\caption{Randomized benchmarking procedure.  (a) State initialization.  Atoms loaded into the lattice are initially in the state $\ket{1,-1}$.  A microwave $\pi$-pulse prepares the atoms in the qubit state $\ket{0}$.  (b) Randomized computational gate sequence.  A series of random microwave $\pi$-pulses (Pauli randomizations, PR) and $\pi/2$-pulses (computational gates, CG) is applied between the qubit states $\ket{0}$ and $\ket{1}$.  A final, calculated $\pi/2$-pulse R returns the qubit to an eigenstate of $\sigma_z$ for measurement.  (c) State read-out.  A microwave $\pi$-pulse maps $\ket{0}$ to $\ket{1,-1}$ in order to read out the relative occupation of the qubit states in a Stern-Gerlach configuration.}
	\label{fig:bm_sequence}
\end{figure}
%

\section{Randomized Benchmarking Experiment}
We perform randomized benchmarking on the atoms confined in the optical lattice using a series of microwave pulses, following the procedure described in Ref.~\cite{knill:benchmarking}.  Our qubit states are the magnetic field-insensitive $\ket{2,0}$ and $\ket{1,0}$ states, defined as $\ket{0}$ and $\ket{1}$, respectively.  Immediately after being loaded into the optical lattice, the confined atoms are in the state $\ket{1,-1}$.  A microwave $\pi$-pulse between $\ket{1,-1}$ and $\ket{0}$ initializes the state of the qubits, as illustrated in Fig.~\ref{fig:bm_sequence}(a).\footnote{The low frequency signal sent to the mixer for the state initialization and state read-out $\pi$-pulses is produced by a commercial DDS signal generator: Novatech 409A.}  A sequence of randomized computational gates is then applied between the $\ket{0}$ and $\ket{1}$ qubit states (Fig.~\ref{fig:bm_sequence}(b)).  A single randomized computational gate is defined as the concatenation of a randomized $\pi$-pulse or identity operation (Pauli randomization, PR) and a $\pi/2$-pulse (computational gate, CG).  An ideal Pauli randomization has the functional form $e^{\pm \ii \sigma_p \pi/2}$, where $\sigma_p$ are the standard Pauli operators with $p = I,x,y,z$.  The sign $\pm$ and Pauli operator $\sigma_p$ are chosen at random with uniform probability at each occurrence in the sequence.  As in previous implementations, the $\sigma_z$ Pauli randomization pulses are implemented only by changing the frame for all subsequent pulses~\cite{knill:benchmarking, ryan:nmr_benchmarking, chow:solid-state_benchmarking}.  Although no pulse is applied for the $\sigma_I$ and $\sigma_z$ operations, the duration of the operation in the sequence is kept the same across all Pauli randomizations.  The form of an ideal computational gate is $e^{\pm \ii \sigma_c \pi/4}$, with the sign $\pm$ and $c = x,y$ each chosen at random with equal probability.  Although some previous implementations of randomized benchmarking have included the change of frame operation ($e^{\pm \ii \sigma_z \pi/4}$) as a computational gate~\cite{ryan:nmr_benchmarking}, here we follow Ref.~\cite{knill:benchmarking, biercuk:q_control, chow:solid-state_benchmarking, chow:optimal_control} and only implement $x,y$ rotations as computational gates.  Following the sequence of randomized computational gates, a final $\pi/2$-pulse with random sign (nested between two more PR pulses) is performed, that is calculated to return the qubit state to an eigenstate of $\sigma_z$ in the absence of errors.  The qubit state is read out by mapping $\ket{0}$ to $\ket{1,-1}$ with a microwave $\pi$-pulse (Fig.~\ref{fig:bm_sequence}(c)), lowering the depth of the optical lattice, and releasing the atoms in a magnetic field gradient.  This Stern-Gerlach type analysis spatially separates the different spin states $\ket{1,-1}$ and $\ket{1}$ $(\equiv \ket{1,0})$ in time-of-flight, allowing us to determine the relative population in each by absorption imaging.

\begin{figure}
	\centering
	\includegraphics[width=1.0\columnwidth,keepaspectratio]{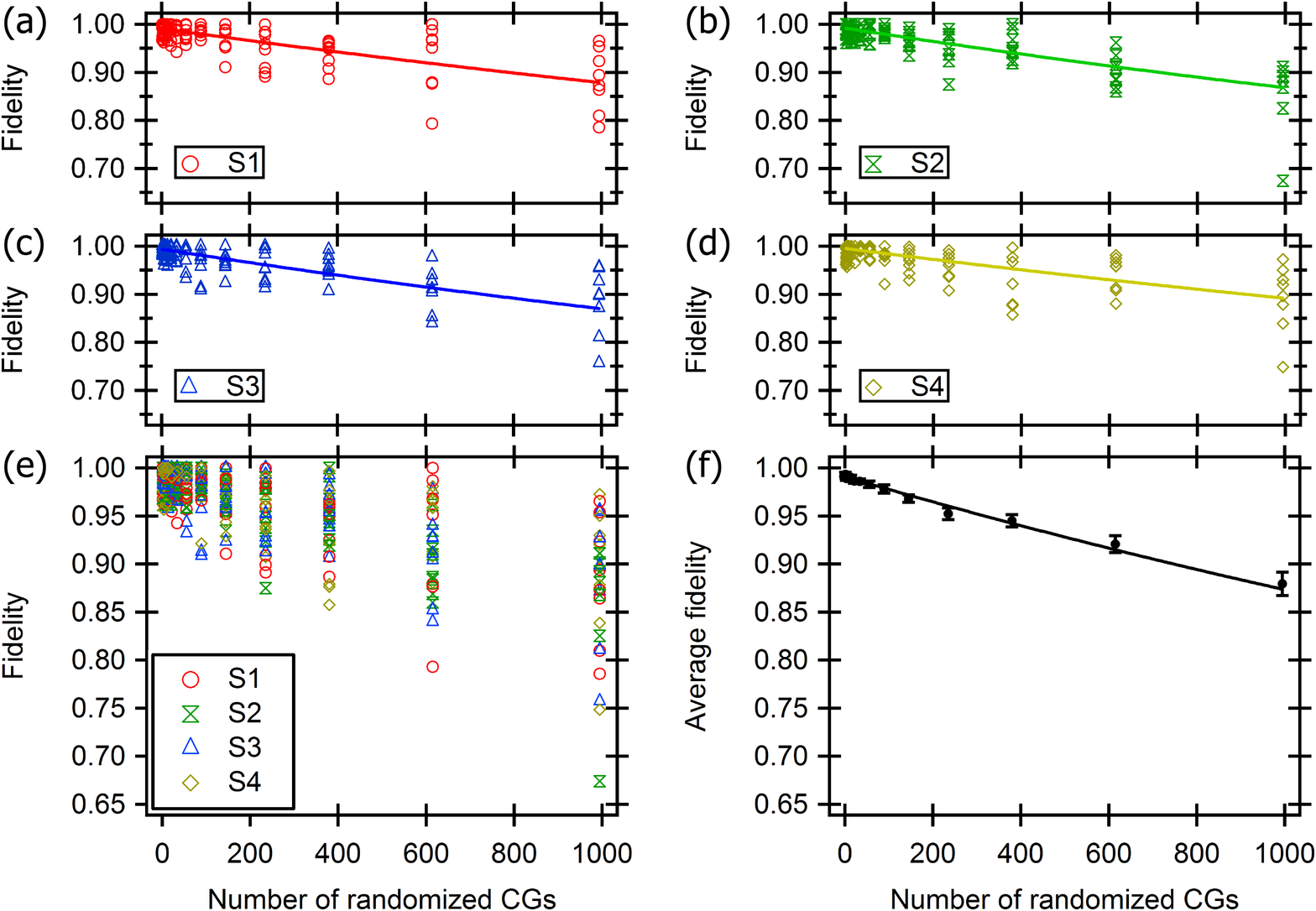}
	\caption{Fidelity of randomized benchmarking sequences versus the number of randomized computational gates (CG).  (a)--(d)  Here, S1, S2, S3, and S4 denote the measurements obtained for the four different random computational sequences, each combined with eight Pauli randomization sequences.  Averaging each over all Pauli randomization sequences and fitting (solid lines) to Eq.~\ref{eq:gate_decay} results in $d = \{2.6(4) \mbox{, } 2.9(3) \mbox{, } 2.9(3) \mbox{, } 2.3(3)\} \times 10^{-4}$ for $\{$S1, S2, S3, S4$\}$.  (e) Measured output state fidelity for all combinations of computational and Pauli randomization sequences.  The scatter in the data suggests that coherent errors may contribute to the decrease in fidelity.  (f)  The output state fidelity averaged over all randomized gate sequences.  The results are fit to Eq.~\ref{eq:gate_decay}, yielding $d_{if} = 1.8(2) \times 10^{-2}$ and $d = 2.7(2) \times 10^{-4}$, for an error probability per randomized computational gate of $\mathcal{E}_g = d/2 = 1.4(1) \times 10^{-4}$.  Error bars are statistical, and represent the standard deviation of the mean.}
	\label{fig:bm_fidelities}
\end{figure}

As the number of randomized computational gates implemented in a sequence is increased, the accumulated gate error reduces the measured single qubit fidelity $\mathcal{F}$ of the output state.  The fidelity is defined as the overlap of the ideal $(\rho_{\mathrm{ideal}})$ and measured $(\rho)$ density matrices, so that $\mathcal{F} = \mbox{tr} \left( \rho_{\mathrm{ideal}} \rho \right)$.  The average output state fidelity $(\overline{\mathcal{F}})$ decays exponentially as~\cite{knill:benchmarking}
\begin{equation}
	\label{eq:gate_decay}
	\overline{\mathcal{F}} = \frac{1}{2} + \frac{1}{2} \left( 1 - d_{if} \right) \left( 1 - d \right)^l
\end{equation}
where $d_{if}$ is the probability of depolarization due to initialization and read-out, $d$ is the average probability of depolarization of a single randomized computational gate (PR and CG), and $l$ is the number of randomized computational gates applied.  The average decrease in fidelity per randomized computational gate (average error per gate) $\mathcal{E}_g$ is related to the probability of depolarization $d$ by~\cite{knill:benchmarking}
\begin{equation}
	\label{eq:error_per_gate}
	\mathcal{E}_g = \frac{d}{2} ,
\end{equation}
giving a fidelity of $1/2$ for a completely depolarized qubit.  Thus, we can apply sequences of varying lengths of randomized computational gates to our single qubits, and fit the decay in the average output state fidelity to Eq.~\ref{eq:gate_decay} to determine $d$ and $\mathcal{E}_g$.

We applied a total of 32 randomized computational gate sequences, derived from 4 random sequences of computational gates and 8 sequences of Pauli randomizations.  Each randomized computational gate sequence was truncated at 15 different lengths \{1, 2, 3, 5, 8, 13, 21, 34, 55, 89, 145, 235, 380, 615, 995\}, for a total of $4 \times 8 \times 15 = 480$ sequences.  The measured output state fidelity for each of these 480 implementations is shown in Fig.~\ref{fig:bm_fidelities}(a)--(e).  The average fidelity of the 32 sequences at each truncation length is given in Fig.~\ref{fig:bm_fidelities}(f), and a weighted fit of this data to Eq.~\ref{eq:gate_decay} yields an average error per gate $\mathcal{E}_g = d/2 = 1.4(1) \times 10^{-4}$.

\section{Evaluating Errors}
We performed several experiments to determine whether control errors or decoherence limit the measured error per computational gate.  The two primary control parameters are the frequency and duration of the applied microwave pulses.  The frequency to drive transitions between the two qubit states is generated by mixing the (high frequency) signal from a microwave signal generator with the (low frequency) output of a custom DDS board.\footnote{Incorporating an AD9951 DDS IC.}  The frequency and phase of the DDS are digitally controlled by a custom field-programmable gate array (FPGA) board,\footnote{Using an Altera Cyclone II IC.} and are adjustable in steps of about 0.1 Hz and 0.02 degrees, respectively.  The duration of each pulse is determined by a TTL output of the FPGA board with 10 ns resolution, which controls an rf switch, resulting in approximately square-shaped pulses.  While the randomized sequences in the benchmarking experiment used at most 995 computational gates (1993 pulses), much longer sequences can be implemented with the present setup.

\begin{figure}
	\centering
	\includegraphics[width=1.0\columnwidth,keepaspectratio]{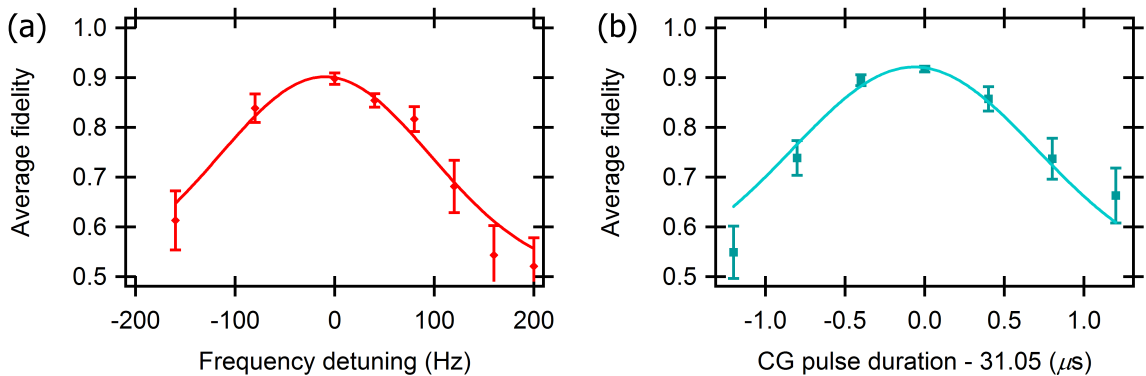}
	\caption{Evaluation of gate control errors using 16 randomized sequences truncated at 500 randomized computational gates.  (a) Fidelity as a function of detuning from the measured $\ket{0}$ to $\ket{1}$ transition.  A gaussian fit to the data gives a peak at $-10(7)$ Hz and width $150(13)$ Hz.  We estimate systematic detuning contributes $< 1 \times 10^{-5}$ to the average error per gate.  (b) The average fidelity as a function of the CG pulse duration, where each PR pulse is twice the CG pulse duration.  Fitting to a gaussian yields a peak at $-0.06(4)$ $\mu$s and a width of $1.1(1)$ $\mu$s.  We thereby approximate that $< 1 \times 10^{-5}$ of the average error per gate results from systematic incorrect pulse duration.  Error bars are the standard deviation of the mean.}
	\label{fig:control_errors}
\end{figure}

The frequency of the qubit transition is acquired by deliberately detuning by about 1 kHz, and performing a detuned Ramsey experiment (two $\pi/2$-pulses separated by a variable time).  A fit to the resulting oscillation, with a delay time as long as 13.5 ms, yields the qubit transition frequency with statistical uncertainty less than 2 Hz.  The measurement also yields an estimate of the unrefocused decay time $T_2^* \approx 25$ ms.

We determine the pulse duration necessary for a $\pi/2$ rotation by varying the time of the applied pulse and fitting to the measured Rabi oscillation.  A typical $\pi/2$ time is 31.05 $\mu$s, with an uncertainty of about 0.01 $\mu$s from the fit.

We characterize the effect of systematic frequency control errors in the benchmarking sequence by performing 16 randomized computational gates sequences (4 CG and 4 PR) at a truncation length of 500 for various detunings.  The CG and PR sequences used here are different than those used in the benchmarking protocol above.  In Fig.~\ref{fig:control_errors}(a) the average output state fidelity at each detuning is plotted.  The output state fidelity is peaked at a detuning near zero ($-10(7)$ Hz), confirming the accuracy of the measured qubit transition frequency.  A gaussian fit yields a width of 150(13) Hz.  Given the uncertainty in the qubit transition frequency above and the deviation from zero of the fitted peak detuning, we estimate the contribution to the average error per gate resulting from systematic detuning as $< 1 \times 10^{-5}$.

A similar evaluation is performed for the duration of the microwave pulses.  Now we measure the average output state fidelity as a function of the pulse durations, with the result shown in Fig.~\ref{fig:control_errors}(b).  Again, the fidelity is peaked near the independently determined $\pi/2$ time of 31.05 $\mu$s.  Since the uncertainty in the nominal pulse duration was only 0.01 $\mu$s, and a gaussian fit to the data in Fig.~\ref{fig:control_errors}(b) gives a width of 1.1(1) $\mu$s and a peak at $-0.06(4)$ $\mu$s, we expect systematic inaccuracies in pulse length contribute $< 1 \times 10^{-5}$ to the average error per gate.

The effect of pulse amplitude noise is estimated by measuring the power spectral density of continuous microwave amplifier output.  Since fluctuations much faster than the pulse length will average to a consistent constant value, and systematic drifts with a time scale much longer than the pulse duration should be manifest in a non-exponential decay of the output state fidelity (absent in the data), we conservatively estimated the potential effect of amplitude noise in the range of about 1 kHz to 100 kHz.  We estimate the total contribution to the average error per gate due to amplitude fluctuations in this range is $< 1 \times 10^{-5}$.

Decoherence of the qubit is another error mechanism, which can arise as a result of inhomogeneity in the system or uncontrolled interactions such as spontaneous scattering of the optical lattice light.  Dephasing due to static inhomogeneity can be counteracted by using a refocusing, or spin-echo, pulse.  While the benchmarking protocol does not explicitly refocus the qubit, some refocusing is expected from the Pauli randomization pulses.

\begin{figure}
	\centering
	\includegraphics[width=1.0\columnwidth,keepaspectratio]{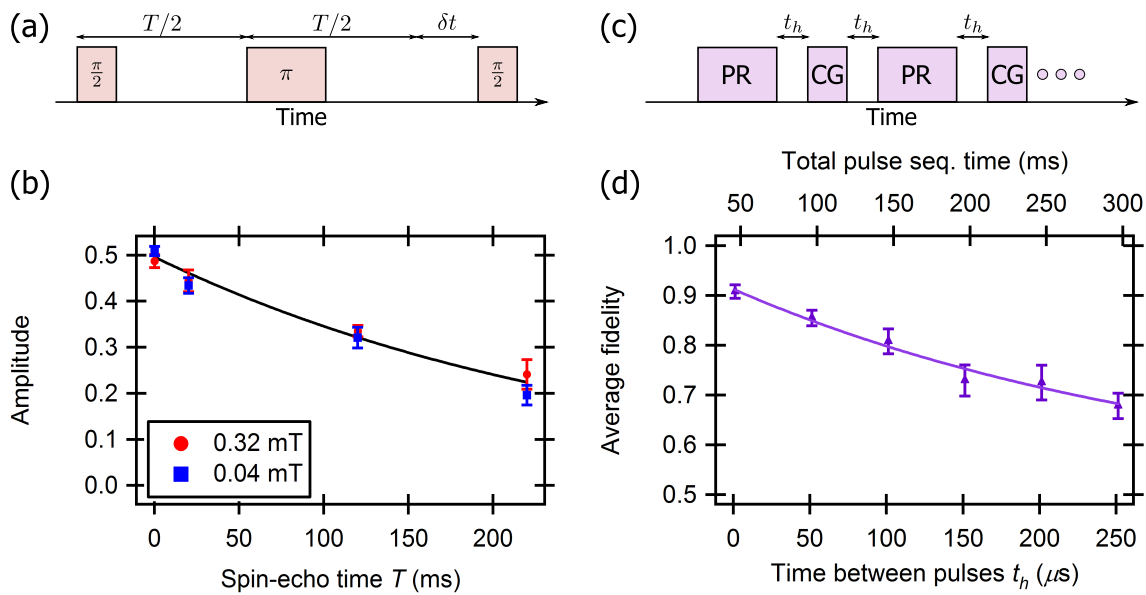}
	\caption{Contribution of decoherence to gate error.  (a) Schematic depiction of determining the decoherence rate in the lattice using a standard detuned spin-echo technique, where for each time $T$ the amplitude of oscillation is obtained by scanning $\delta t$ over about 1.5 ms.  (b) Measured decay in the amplitude of the detuned spin-echo signal as a function of the time $T$, yielding an average decoherence time $T_2 \approx 0.28(2)$~s (exponential fit).  (c) Experiment schematic for measuring the contribution of the duration of the benchmarking sequence to the average error per gate.  PR is a Pauli randomization ($\pi$) pulse; CG is a computational gate ($\pi/2$) pulse.  (d) Average fidelity as a function of the time between pulses in the benchmarking sequence, $t_h$, at a truncation length of 500 randomized computational gates (1003 pulses).  An exponential fit to the average fidelity versus total sequence time yields a decay constant of 0.31(4) s, consistent with $T_2$.  Uncertainties are the statistical standard error.}
	\label{fig:decoherence}
\end{figure}

We measure the decoherence time $T_2$ with a detuned spin-echo experiment, as illustrated in Fig.~\ref{fig:decoherence}(a).  The decay time is measured at two different magnitudes of the external magnetic field with similar results (average decay time $T_2 \approx 0.28(2)$ s), indicating that the increased field sensitivity (from 4 kHz/mT to 37 kHz/mT) of the qubit states at the larger magnetic field does not limit the decoherence rate (Fig.~\ref{fig:decoherence}(b)).  The data presented in Fig.~\ref{fig:bm_fidelities} and~\ref{fig:control_errors} was taken at 0.32 mT.

A more direct measure of decoherence in the presence of the partial refocusing from the benchmarking sequence is achieved by simply extending the total time over which the sequence of randomized computational gates is applied.  As displayed in Fig.~\ref{fig:decoherence}(c), we accomplish this by adding an additional time $t_h$ between subsequent pulses in the benchmarking sequence.  Similar to the control error evaluation, a total of 16 randomized computational gate sequences (4 PR, 4 CG) at a truncation of 500 are applied to the atoms, with the average output fidelity for each value of $t_h$ shown in Fig.~\ref{fig:decoherence}(d).  A simple exponential fit to the average fidelity as a function of the total pulse sequence time gives a decay constant of 0.31(4) s, consistent with the spin-echo measured $T_2$ average value of 0.28(2) s.  Given that a single randomized computational gate (PR and CG) is typically 94 $\mu$s, a $T_2$ decay during this time would result in an error per gate of about $1.6(3) \times 10^{-4}$.  Thus, the primary contributor to the error in a single qubit gate appears to be the $T_2$ coherence time.

A simple Monte Carlo simulation of the benchmarking sequences is used to model the effective refocusing of the qubit.  Although the benchmarking sequences do not explicitly refocus the qubit, the simulation indicates that the qubit is almost completely refocused for a sufficiently long truncation of the benchmarking sequence.  From the Monte Carlo simulation, we estimate the contribution of inhomogeneity across the atom cloud to the error per gate to be $< 1 \times 10^{-5}$, confirming that the dominate contribution to the error is the decoherence characterized by $T_2$.  

\section{Outlook}
Randomized benchmarking on atomic qubits in an optical lattice reinforces the viability of this system for applications in quantum information.  A series of experiments demonstrates that we have precise control of the single qubit gate operations, without requiring composite pulse sequences or employing pulse shaping techniques.  Although these methods have been shown to reduce the influence of control errors~\cite{ryan:nmr_benchmarking, chow:optimal_control, cummins:composite_pulses, timoney:err-resist_gates, rakreungdet:uw_control} and may be useful in future implementations, the gate operations are currently limited by the coherence time ($T_2$) of the system, which is probably constrained by the optical lattice.

Spontaneous Raman scattering from the lattice is estimated from experimental parameters to occur at a rate of about 0.2~s$^{-1}$.  Assuming that a single Raman scattering event depolarizes the qubit while Rayleigh scattering retains the coherence~\cite{cline:spin_relax, ozeri:coherence_scattering}, spontaneous scattering then results in an error of about $1 \times 10^{-5}$ per randomized computational gate.  The fidelity of the single qubit operations is probably limited by other aspects of the current optical lattice arrangement, such as time-dependent frequency shifts resulting from lattice beam intensity fluctuations.  The scatter in Fig.~\ref{fig:bm_fidelities}(e) could be a result of these coherent errors.  The operation of the quantum gate may be improved by better intensity stabilization of the lattice beams, or by operating the lattice in a regime that is insensitive to differential scalar light shifts~\cite{lundblad:diff_light_cancel, derevianko:doubly-magic_lattice, dudin:light_store_mag_lattice, chicireanu:diff_light_shifts}.

Advanced applications in quantum information will require individual qubit addressing.  The microwave gate operations demonstrated here can be made to address single lattice sites by integrating high numerical aperture optics in the system to allow for tightly focused addressing beams~\cite{zhang:single_qubits_lattice}.  In this configuration, the average error per gate resulting from both spontaneous scattering in the presence of the addressing beam and off-resonant microwave transitions in neighboring lattice sites can be $< 1 \times 10^{-4}$~\cite{zhang:single_qubits_lattice, lundblad:field-sensitive_addressing}.\footnote{In this estimate we assume lattice beams of wavelength 1.064 $\mu$m, an addressing beam operating at 789.5 nm with beam waist 1 $\mu$m, and gaussian-shaped microwave pulses.}  Thus, the combined addressing beam and optical lattice system should be able to utilize the high fidelity quantum gate operations presented here in a scalable architecture, facilitating more advanced quantum information protocols.

\ack
We thank E. Knill, C. A. Ryan, and A. Meier for useful discussions regarding the randomized benchmarking procedure, and E. Knill, T. M. Hanna, and A. Meier for comments on the manuscript.  Specific product citations are for the purpose of clarification only, and are not an endorsement by the authors, JQI or NIST.  S.O. and K.D.N. acknowledge support from the National Research Council (NRC) Research Associateship program.  This work was partially supported by the DARPA QUEST program.  This paper is a contribution of NIST and is not subject to U.S. copyright.

\section*{References}


\begin{thebibliography}{10}

\bibitem{ladd:qc_expt_review}
T.~D. Ladd, F.~Jelezko, R.~Laflamme, Y.~Nakamura, C.~Monroe, and J.~L.
  {O'Brien}.
\newblock Quantum computers.
\newblock {\em Nature}, {\bf 464}:45, 2010.

\bibitem{kuah:prep_qpt}
A.~Kuah, K.~Modi, C.~A. {Rodr\'{i}guez-Rosario}, and E.~C.~G. Sudarshan.
\newblock How state preparation can affect a quantum experiment: Quantum
  process tomography for open systems.
\newblock {\em Phys. Rev. A}, {\bf 76}:042113, 2007.

\bibitem{mohseni:qpt}
M.~Mohseni, A.~T. Rezakhani, and D.~A. Lidar.
\newblock Quantum-process tomography: Resource analysis of different
  strategies.
\newblock {\em Phys. Rev. A}, {\bf 77}:032322, 2008.

\bibitem{knill:benchmarking}
E.~Knill et~al.
\newblock Randomized benchmarking of quantum gates.
\newblock {\em Phys. Rev. A}, {\bf 77}:012307, 2008.

\bibitem{biercuk:q_control}
M.~J. Biercuk, H.~Uys, A.~P. Vandevender, N.~Shiga, W.~M. Itano, and J.~J.
  Bollinger.
\newblock High-fidelity quantum control using ion crystals in a Penning trap.
\newblock {\em Quant. Inf. Comp.}, {\bf 9}:920, 2009.

\bibitem{ryan:nmr_benchmarking}
C.~A. Ryan, M.~Laforest, and R.~Laflamme.
\newblock Randomized benchmarking of single- and multi-qubit control in
  liquid-state NMR quantum information processing.
\newblock {\em New J. Phys.}, {\bf 11}:013034, 2009.

\bibitem{chow:solid-state_benchmarking}
J.~M. Chow et~al.
\newblock Randomized Benchmarking and Process Tomography for Gate Errors in a
  Solid-State Qubit.
\newblock {\em Phys. Rev. Lett.}, {\bf 102}:090502, 2009.

\bibitem{chow:optimal_control}
J.~M. Chow, L.~DiCarlo, J.~M. Gambetta, F.~Motzoi, L.~Frunzio, S.~M. Girvin,
  and R.~J. Schoelkopf.
\newblock Implementing optimal control pulse shaping for improved single-qubit
  gates.
\newblock 2010.
\newblock arXiv:1005.1279v1.

\bibitem{phillips:zeeman_slower}
W.~D. Phillips and H.~Metcalf.
\newblock Laser Deceleration of an Atomic Beam.
\newblock {\em Phys. Rev. Lett.}, {\bf 48}:596, 1982.

\bibitem{raab:mot}
E.~L. Raab, M.~Prentiss, A.~Cable, S.~Chu, and D.~E. Pritchard.
\newblock Trapping of Neutral Sodium Atoms with Radiation Pressure.
\newblock {\em Phys. Rev. Lett.}, {\bf 59}:2631, 1987.

\bibitem{pritchard:rf_evap}
D.~E. Pritchard, K.~Helmerson, and A.~G. Martin.
\newblock Atom Traps.
\newblock In {\em Proceedings of the Eleventh International Conference on
  Atomic Physics}, page 179. World Scientific, Singapore, 1989.

\bibitem{lin:quad_dipole_bec}
Y.-J. Lin, A.~R. Perry, R.~L. Compton, I.~B. Spielman, and J.~V. Porto.
\newblock Rapid production of ${}^{87}$Rb Bose-Einstein condensates in a
  combined magnetic and optical potential.
\newblock {\em Phys. Rev. A}, {\bf 79}:063631, 2009.

\bibitem{sebby-strabley:double-well_lattice}
J.~{Sebby-Strabley}, M.~Anderlini, P.~S. Jessen, and J.~V. Porto.
\newblock Lattice of double wells for manipulating pairs of cold atoms.
\newblock {\em Phys. Rev. A}, {\bf 73}:033605, 2006.

\bibitem{ovchinnikov:atom_diffraction}
Y.~B. Ovchinnikov, J.~H. M{\"u}ller, M.~R. Doery, E.~J.~D. Vredenbregt,
  K.~Helmerson, S.~L. Rolston, and W.~D. Phillips.
\newblock Diffraction of a Released Bose-Einstein Condensate by a Pulsed
  Standing Light Wave.
\newblock {\em Phys. Rev. Lett.}, {\bf 83}:284, 1999.

\bibitem{campbell:mott_shells}
G.~K. Campbell, J.~Mun, M.~Boyd, P.~Medley, A.~E. Leanhardt, L.~G. Marcassa,
  D.~E. Pritchard, and W.~Ketterle.
\newblock Imaging the Mott Insulator Shells by Using Atomic Clock Shifts.
\newblock {\em Science}, {\bf 313}:649, 2006.

\bibitem{cummins:composite_pulses}
H.~K. Cummins, G.~Llewellyn, and J.~A. Jones.
\newblock Tackling systematic errors in quantum logic gates with composite
  rotations.
\newblock {\em Phys. Rev. A}, {\bf 67}:042308, 2003.

\bibitem{timoney:err-resist_gates}
N.~Timoney, V.~Elman, S.~Glaser, C.~Weiss, M.~Johanning, W.~Neuhauser, and
  C.~Wunderlich.
\newblock Error-resistant single-qubit gates with trapped ions.
\newblock {\em Phys. Rev. A}, {\bf 77}:052334, 2008.

\bibitem{rakreungdet:uw_control}
W.~Rakreungdet, J.~H. Lee, K.~F. Lee, B.~E. Mischuck, E.~Montano, and P.~S.
  Jessen.
\newblock Accurate microwave control and real-time diagnostics of neutral-atom
  qubits.
\newblock {\em Phys. Rev. A}, {\bf 79}:022316, 2009.

\bibitem{cline:spin_relax}
R.~A. Cline, J.~D. Miller, M.~R. Matthews, and D.~J. Heinzen.
\newblock Spin relaxation of optically trapped atoms by light scattering.
\newblock {\em Op. Lett.}, {\bf 19}:207, 1994.

\bibitem{ozeri:coherence_scattering}
R.~Ozeri et~al.
\newblock Hyperfine Coherence in the Presence of Spontaneous Photon Scattering.
\newblock {\em Phys. Rev. Lett.}, {\bf 95}:030403, 2005.

\bibitem{lundblad:diff_light_cancel}
N.~Lundblad, M.~Schlosser, and J.~V. Porto.
\newblock Experimental observation of magic-wavelength behavior of ${}^{87}$Rb
  atoms in an optical lattice.
\newblock {\em Phys. Rev. A}, {\bf 81}:031611(R), 2010.

\bibitem{derevianko:doubly-magic_lattice}
A.~Derevianko.
\newblock {``Doubly Magic''} Conditions in Magic-Wavelength Trapping of
  Ultracold Alkali-Metal Atoms.
\newblock {\em Phys. Rev. Lett.}, {\bf 105}:033002, 2010.

\bibitem{dudin:light_store_mag_lattice}
Y.~O. Dudin, R.~Zhao, T.~A.~B. Kennedy, and A.~Kuzmich.
\newblock Light storage in a magnetically-dressed optical lattice.
\newblock {\em Phys. Rev. A}, {\bf 81}:041805(R), 2010.

\bibitem{chicireanu:diff_light_shifts}
R.~Chicireanu, K.~D. Nelson, S.~Olmschenk, N.~Lundblad, A.~Derevianko, and
  J.~V. Porto.
\newblock Differential Light Shift Cancellation in a Magnetic-Field-Insensitive
  Transition of ${}^{87}$Rb.
\newblock 2010.
\newblock arXiv:1010.1520v1.

\bibitem{zhang:single_qubits_lattice}
C.~Zhang, S.~L. Rolston, and S.~{Das Sarma}.
\newblock Manipulation of single neutral atoms in optical lattices.
\newblock {\em Phys. Rev. A}, {\bf 74}:042316, 2006.

\bibitem{lundblad:field-sensitive_addressing}
N.~Lundblad, J.~M. Obrecht, I.~B. Spielman, and J.~V. Porto.
\newblock Field-sensitive addressing and control of field-insensitive
  neutral-atom qubits.
\newblock {\em Nature Physics}, {\bf 5}:575, 2009.

\end{thebibliography}

\end{document}